\documentclass[12pt]{iopart}

 \usepackage{epsfig,bm}
 \def\dis{distribution}
\def\pt{$p_T$}
\def\tr{$p_T^{\rm trig}$}
\def\t{p_T^{\rm trig}}
\def\as{$p_T^{\rm assoc}$}
\def\a{p_T^{\rm assoc}}
\def\om{$\Omega$}
\def\ass{associated }
\def\pa{particles }
\begin{document}

\title[The Omega Puzzle]{The Omega Puzzle}

\author{Rudolph C. Hwa}

\address{Institute of Theoretical Science and Department of Physics, University of Oregon, Eugene, OR 97403-5203, USA}
\ead{hwa@uoregon.edu}
\begin{abstract}
Recent data on particles associated  with \om\ trigger suggest that such events have jet structure. However, the exponential \pt\ \dis\ of \om\ suggests that no jets are involved. We resolve this puzzle by suggesting that the \om\ is produced by the recombination of the enhanced thermal partons in the ridge that can have other hadrons in association with the trigger, all having exponential \pt\ \dis s.
\end{abstract}

At the plenary sessions of  this meeting it was shown twice that the STAR data on $\Omega$ production at $2.5<\t<4.5$ GeV/c have associated particles above background for $1.5<\a<\t$ \cite{1,2}. More details were given by Bielcikova \cite{3} showing a peak in $|\Delta\phi|<0.5$ that falsifies the prediction that \om\ trigger at high \pt\ should not be accompanied by associated particles on the near or far sides \cite{4}. The data created a puzzle that needed to be addressed urgently. We present here the solution of that puzzle, instead of the original material prepared for the talk whose title appeared in the program. The replaced subjects are on the large $p/\pi$ ratios in particles produced  (a) at intermediate \pt\ in the forward region at RHIC \cite{5}, and (b) in the interval $10<p_T<20$ GeV/c in the mid-rapidity region at LHC \cite{6}. In both of those regions no \ass \pa above background are expected. Readers are referred to \cite{5,6} for details. We restrict our attention here to the \om\ puzzle only.

In \cite{4} on the strange-particle production at intermediate \pt, it is shown that $s$ quark shower parton is suppressed, and that \om\ produced in the range $1<p_T<6$ GeV/c at $\sqrt s=200$ GeV in central Au+Au collisions can be understood as the result of the recombination of three thermal partons with no involvement of shower partons. It is then predicted that there would be no particles \ass with \om\ above background. The recent data on the \om-triggered events \cite{1,2,3} prove that the prediction is wrong, although the assertion of  the absence of shower partons in \om\ is unchallenged. We thus have two seemingly contradictory phenomena that form the puzzle: (1) The \om\ spectrum is exponential up to $p_T\sim 6$ GeV/c, and (2) the \om-triggered events in the range $2.5<\t<4.5$ GeV/c have associated hadrons in the range $1.5<\a<\t$. The two phenomena are depicted by the two panels in Fig.\ 1.
It is the aim of this talk to offer a solution to this puzzle.

The implications of these two phenomena are as follows. Item (1) means that there is no contribution from hard scattering, which has the characteristic that it behaves in \pt\ as a power law.  Without that, there is no jet. However, item (2) means that there is jet structure. So is there or is there not a jet? The resolution of the puzzle resides in the recognition of {\it phantom jet}.

\begin{figure}[h]
\null\vspace*{0.4cm}
{\hspace*{-.3cm}
\includegraphics[scale=0.35]{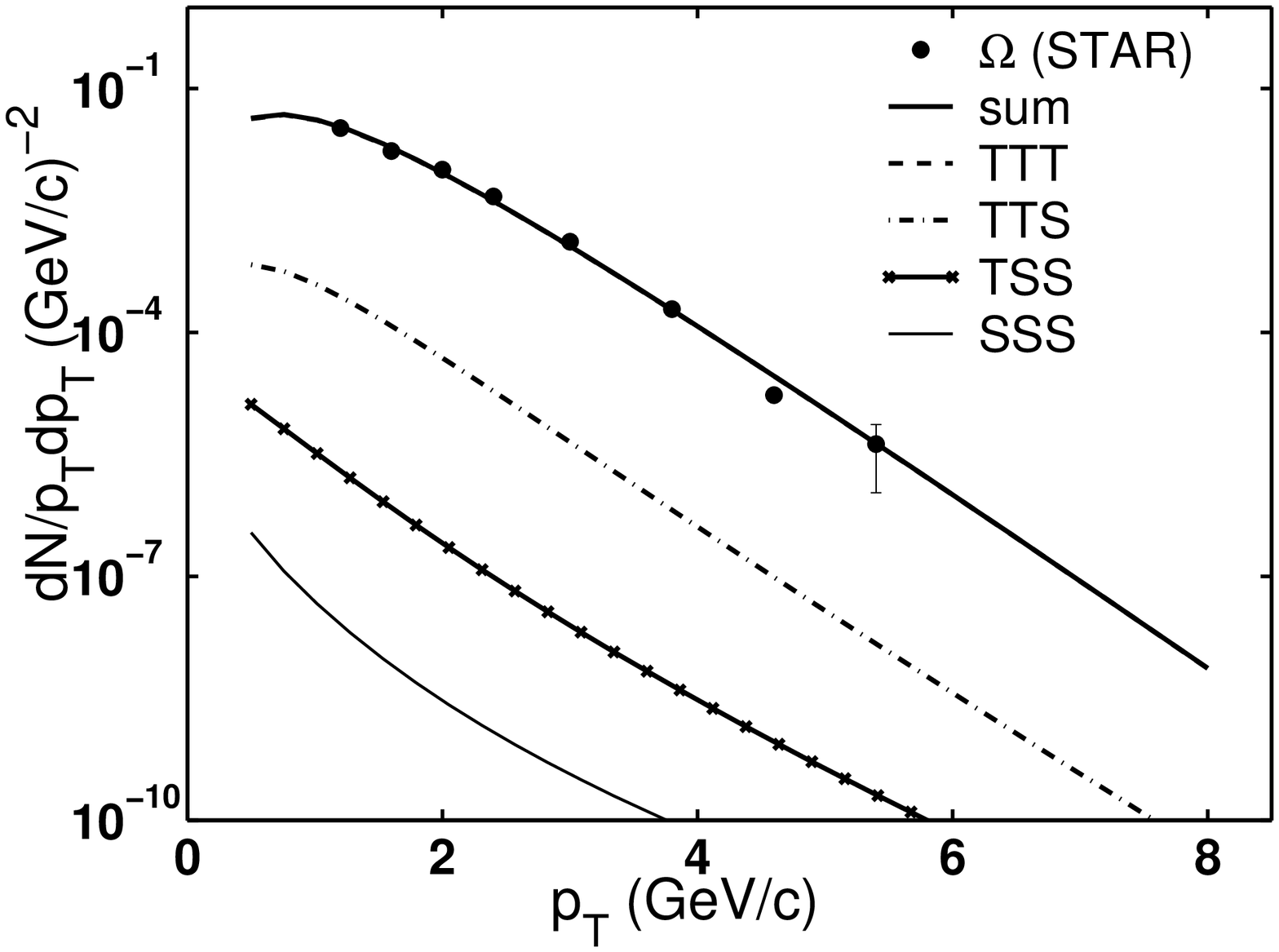}}
{\hspace*{.1cm}\vspace*{-8.3cm}
\includegraphics[scale=0.4]{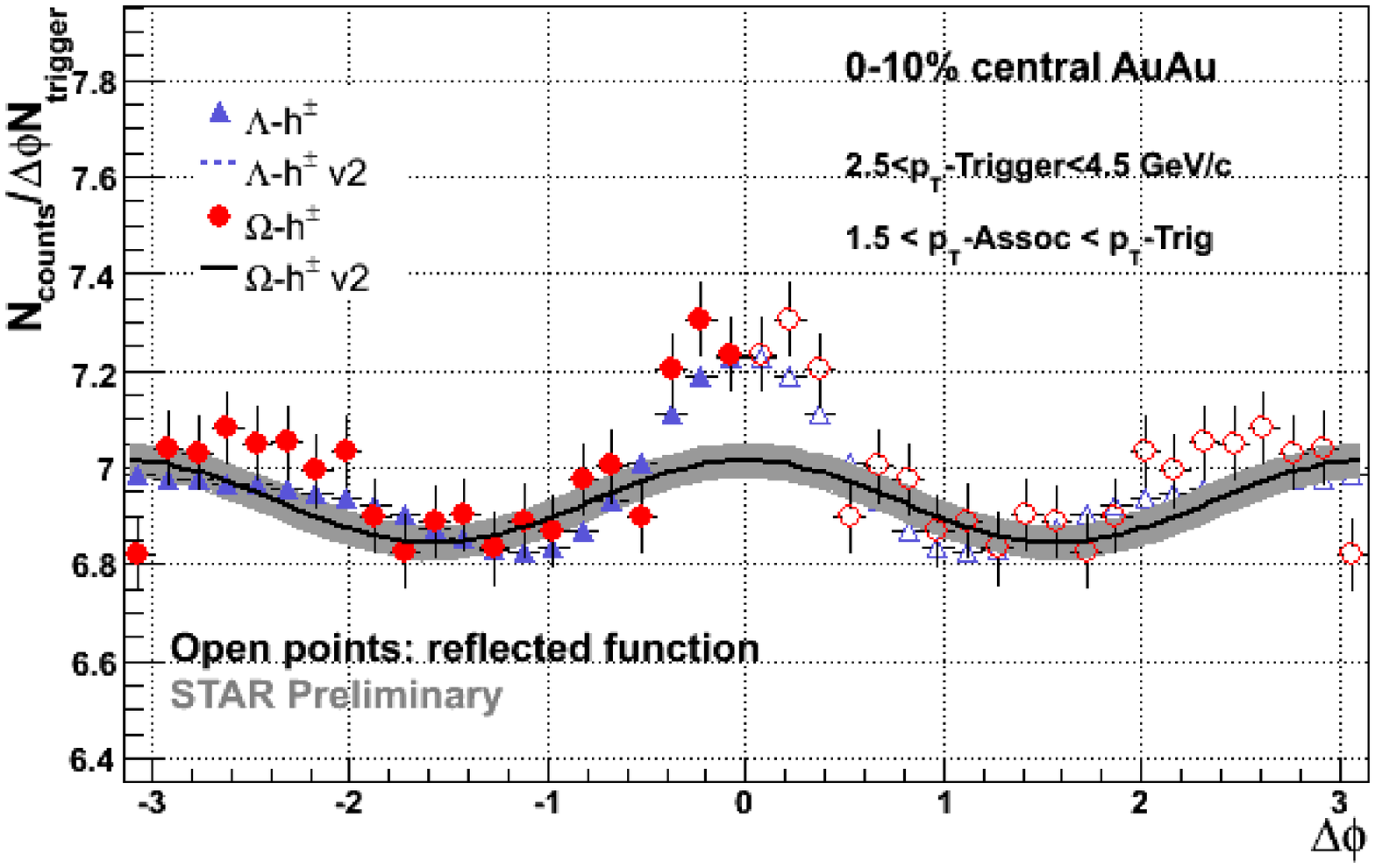}}
\vspace*{8cm}
\caption{ (Left panel) Transverse momentum distribution of \om\ in central Au+Au collisions. The TTT contribution is covered up by the sum in solid line. Data are from \cite{7}; the figure is from \cite{4}. (Right panel) Associated particle distribution of \om-triggered events in red points \cite{1,2,3}.}
\end{figure}

Before going any further, it is helpful to review first jet phenomenology, as discussed by Calderon \cite{8}. The jet structure on the near side has a peak and a ridge. They are both visible when projected on $\Delta\eta$, since the ridge is a flat pedestal on top of which sits the peak. When projected on $\Delta\phi$ the ridge is not distinguishable from the peak because the ridge has a limited range in $\Delta\phi$. It should be noted that the properties of these projections have been discussed in the recombination model in \cite{9}, to which we shall return below. How the peak and ridge depend on $N_{\rm part}$, \tr\ and \as\ was shown by Putschke  \cite{10}. STAR has chosen to refer to the peak  as ``Jet", which can be misleading, since the ridge is also a part of the jet in the more inclusive sense. Despite possible confusion we shall adhere to the nomenclature used in the experimental presentations where the peak was referred to as ``Jet" (with capital J).

The result of the jet structure analysis on unidentified \pa with $3<\t<4$ GeV/c and $\a>2$ GeV/c is that the yield of Jet+Ridge($\Delta\phi$) increases with $N_{\rm part}$, while that for Jet($\Delta\phi$) or Jet($\Delta\eta$) is essentially independent of $N_{\rm part}$. That  is a very striking result, since we know that the $p/\pi$ ratio at those \pt\ values depends on centrality, increasing from $~0.1$ in $pp$ collision to $\sim 1$ in central Au+Au collisions.  The $J+R$ yield being about 3 times the $J$ yield at 0-10\% centrality suggests that the ridge $R$ not only is high in yield but also contributes significantly to the $p/\pi$ ratio. Putschke further showed that at 0-10\% centrality the behavior of the $J$ yield in \as\ depends on \tr\ but that of the $R$ yield in \as\ does not \cite{10}. For $\a>2.3$ GeV/c and $\t>4$ GeV/c, it is found that $J/R\ ^>_\sim\ 1$.

Pushing \as\ to lower values, Bielcikova showed that the $J/R$ ratio becomes as low as 0.1-0.15 at $\a\sim $1.2 GeV/c for unidentified hadrons, and that it is even lower $(<0.1)$  for $\Lambda$-triggered events \cite{3,11}. We expect that for \om-triggered events the $J/R$ ratio would be even smaller, perhaps $\sim 0.01$, in which case the peak would be hardly visible in the $\Delta\eta$ projection. That is the scenario for which the notion of phantom jet is appropriate in referring to a $\Delta\eta$ \dis\ that has essentially only a ridge. A phantom jet is a jet, except that its manifestation at low \as\ lacks the usual peak that is regarded as the Jet.
It should be emphasized that how an ordinary jet may appear as a phantom jet depends on \tr\ and \as .

In \cite{9} we have presented an interpretation of the ridge (called pedestal there) and explained why it appears visibly in the $\Delta\eta$ \dis , but not in $\Delta\phi$. Its dynamical origin is a hard parton traversing a portion of the medium near the surface (how near the surface depends on how high \tr\ is); the energy loss to the medium enhances the thermal partons locally, which in turn hadronize by thermal-thermal recombination resulting in the ridge. Thus the ridge is a part of the jet produced by a high-\pt\ parton, but a jet does not necessarily have a ridge, as when \tr\ is so high as to select events in which the hard scattering occurs very near the surface. The peak above the ridge is a more direct consequence of hard scattering and is reproduced by the recombination of  shower and thermal partons \cite{9}. The \pt\ \dis\ of a pion that gives rise to the peak bends up from the exponential behavior of thermal-thermal recombination toward  a power-law behavior \cite{12}.

The events with the \om\ trigger are at the low end of \tr\ between 2.5 and 4.5 GeV/c. If one adopts the conventional presumption that a trigger particle is the fragment of a hard parton, then one would expect that there are, as usual, associated particles on the near side. If that is true, then the \om\ spectrum would be power-law behaved, and $J/R$ would not be infinitesimal. Currently, there are not enough \om-triggered events to separate the Jet from the Ridge. But we do know that the \pt\ \dis\ of \om\ is essentially exponential \cite{7} (see Fig.\ 1). Since the  $s$ quarks in hard scattering are suppressed, no shower partons contribute to  \om\ \cite{4}. The ridge being formed from the enhanced thermal partons contains $s$ quarks. The recombination of 3 $s$-quarks in the ridge can form the trigger \om . Since \tr\ is $\ge$ 2.5 GeV/c, the $s$ quarks need only have $p_T\approx 0.8$ GeV/c each to form the \om , which does not strain the thermal source. The light quarks in the ridge can easily form other hadrons with $\a>1.5$ GeV/c.

The recognition of \om\ being formed from the ridge then solves the puzzle. Being of thermal origin, its \pt\ \dis\ is exponential. On the other hand, the ridge being above the statistical background can be the source of associated \pa  seen in the $\Delta\phi$ \dis\ \cite{3}, as has been obtained in the pion-triggered near-side \dis\ studied in the recombination model \cite{9}. The excess of hadrons observed above the statistical background in the right panel of Fig.\ 1 is entirely in the ridge.

To have the ridge without the Jet (peak) is my conjecture based on a reasonable extrapolation from the observation that $J/R$ for $\Lambda$ trigger is $<0.1$ \cite{3,11}. The solution of the \om\ puzzle does not depend crucially on the absence of the peak (consisting of unidentified hadrons), but does depend on the presence of a large ridge when \as\ is extended to as low as 1.5 GeV/c. The implication of phantom jet is quite appropriate, since the \om\ production is actually due to a very hard scattering that gives rise to a sizable ridge, which can be accompanied by a Jet that can be seen if  \as\ were higher. But that Jet is not seen when \as\ is pushed down to 1.5 GeV/c for the purpose of increasing the statistics of observing the associated particles of the \om\ trigger. At \as\ as low as that, the ridge overwhelms the Jet. In other words the phantom jet   is not seen as a Jet; its shadow is the ridge that is seen. 

One may ask how much of the calculation done in \cite{4} has been affected by the solution of the \om\ puzzle. None at all. In all our study of the spectra at intermediate \pt\ we have always determined the thermal parton distribution from the data by fitting the soft hadron \dis s at $p_T<2$ GeV/c, since our investigations never attempted to solve the low-\pt\ problem \cite{13}. In \cite{4} we have found for $\phi$ production that $T_s=0.392$ GeV, which is the same ``temperature" used to reproduce the \om\ \dis. That value of $T_s$ is $\sim 20$\% higher than the value $T=0.317$ GeV determined in the study of pion and proton production \cite{13}. The difference can partially be attributed to the flow effect of the more massive $s$ quarks, but now should also be recognized as being partially due to the enhanced thermal \dis\ of the ridge. The assertion made in \cite{4} about no associated particles in $\phi$- or \om-triggered events was based solely on the trigger being formed from the thermal partons. Our error was in identifying the thermal partons as being in the statistical background. Now, we identify them as being in the ridge that is above the background. That is a change in the qualitative prediction that involves no change in the quantitative calculation. Thus the solution of the \om\ puzzle does not invalidate any of the calculated results in \cite{4}.

To test our solution of the \om\ puzzle, the following predictions can be checked. The ridge should be seen in $\Delta\eta$. The hadrons associated with \om\ should have exponential behavior in \pt . Finally, among the associated particles the $p/\pi$ ratio should be high, since both hadrons are formed by recombination from the thermal partons in the ridge.

This work was supported  in part,  by the U.\ S.\ Department of Energy under Grant No. DE-FG02-96ER40972.

\end{document}